\documentclass[12pt, peerreview , onecolumn]{IEEEtran}

\RequirePackage[normalem]{ulem} %DIF PREAMBLE
\RequirePackage{color}\definecolor{RED}{rgb}{1,0,0}\definecolor{BLUE}{rgb}{0,0,1} %DIF PREAMBLE
\RequirePackage{color}
\usepackage{tikz}
\usepackage[sumlimits,intlimits]{amsmath}
\usepackage{amsmath,amsfonts,amssymb}%
\usepackage{bm}%
\usepackage{graphicx,graphics}%
\usepackage{cases}%
\usepackage[noadjust]{cite}%
\usepackage{color}%
\usepackage{cite,url}
\usepackage{verbatim}
\usepackage{enumerate}
\usepackage{balance}
\usepackage{multirow}
\usepackage{stfloats}
\usepackage{subfigure}
\usepackage{xspace}
\usepackage{smartref}
\usepackage[lined,algoruled,linesnumbered]{algorithm2e}

\ifCLASSOPTIONpeerreview

\fi

\begin{document}

\title{\vspace*{-10pt}Joint Phase Noise Estimation and Data Detection in Coded MIMO Systems}

%\author{\vspace*{0pt}
%\authorblockN{Arif Onder Isikman$^\dag$, Hani Mehrpouyan$^\ddag$, Ali A. Nasir$^\S$ and Alexandre Graell i Amat$^\dag$.\vspace*{-0pt}}
%\IEEEauthorblockA{\\$^\dag$ Chalmers University of Technology, Sweden.\\
%$^\ddag$ Department of ECE\&CS, California State University, Bakersfield, USA.\\
%$^\S$ Australian National University, Australia.\\
%emails: isikman@student.chalmers.se, hani.mehr@ieee.org, ali.nasir@anu.edu.au, alexandre.graell@chalmers.se
%\vspace{-10pt}
%}}

\author{\vspace*{-0pt}
\authorblockN{Arif O. Isikman$^\dag$,
                \hspace{0pt}Hani Mehrpouyan$^\ddag$,~\IEEEmembership{Member, IEEE,}
                Ali A. Nasir$^\sharp$,~\IEEEmembership{Member, IEEE},
                Alexander G. Amat{$^\dag$},~\IEEEmembership{Senior Member, IEEE},
                and Rodney. A. Kennedy$^\sharp$,~\IEEEmembership{Fellow, IEEE}.
}\\
\vspace{+0pt}
$^\dag$Communications Systems Group at Chalmers University of Technology, Sweden,
\newline
$^\ddag$Department of ECE at California State University, Bakersfield, USA,
\newline
$^\sharp$The Research School of Engineering at the Australian National University, Australia.
\newline
\{Emails: isikman@student.chalmers.se, hani.mehr@ieee.org, ali.nasir@anu.edu.au, alexandre.graell@chalmers.se, and rodney.kennedy@anu.edu.au\}
}
\maketitle
{\let\thefootnote\relax\footnotetext{{This work is part of the first author's Master dissertation, which was completed
at Chalmers University of Technology and Ericsson AB \cite{Arif-2012}. This research was supported by Swedish research foundation Vinnova and the Australian Research Council's Discovery Project funding scheme (project number DP110102548).  \vspace{-0pt}}} }

\vspace{-12pt}

\begin{abstract}
In this paper, the problem of joint oscillator \emph{phase noise (PHN)} estimation and data detection for \emph{multi-input multi-output (MIMO)} systems using \emph{bit-interleaved coded modulation (BICM)} is analyzed. A new MIMO receiver that iterates between the estimator and the detector, based on the \emph{expectation-maximization (EM)} framework, is proposed. It is shown that at high signal-to-noise ratios, a \emph{maximum a posteriori estimator (MAP)} can be used to carry out the maximization step of the EM algorithm. Moreover, to reduce the computational complexity of the proposed EM algorithm, a soft decision-directed \emph{extended Kalman filter-smoother (EKFS)} is applied instead of the MAP estimator to track the PHN parameters. Numerical results show that by combining the proposed EKFS based approach with an iterative detector that employs \emph{low density parity check (LDPC)} codes, PHN can be accurately tracked. Simulations also demonstrate that compared to existing algorithms, the proposed iterative receiver can significantly enhance the performance of MIMO systems in the presence of PHN.

%to calculate the marginal a posteriori probabilities of the transmitted symbols. Numerical investigations show that for a wide range of phase noise variances the error performance of the proposed EM based algorithm improves at every iteration.
\end{abstract}

\vspace*{-0pt}

%\newpage
\vspace{-12pt}
\section{Introduction}
%Noisy oscillators introduce time varying PHN to digital communication systems. Since the Lorentzian spectrum of oscillators has a slope of $1/f^2$ at high carrier frequencies, the time varying PHN process is commonly modeled as a Wiener process~\cite{Demir}.

\vspace{-10pt}
\subsection{Motivation and Literature Survey}
It is well-know that \emph{multi-input multi-output (MIMO)} technology allows for more efficient use of the available spectrum~\cite{Telatar99}. To this end, \emph{bit-interleaved-coded-modulation (BICM)} is a popular scheme that enables communication systems to fully exploit the spectrum efficiency promised by MIMO technology~\cite{Duman}. However, the performance of MIMO systems degrades dramatically in the presence of synchronization errors. In fact, one of the main limiting factors in the deployment of MIMO systems in microwave links, e.g., for establishing the backhaul link, is \emph{phase noise (PHN)}~\cite{Wells-2010}.

Analogous to other circuits, the oscillator circuitry is affected by thermal noise. As such, the output of practical oscillators is not perfectly periodic and is affected by PHN. PHN interacts with the transmitted symbols in a non-linear manner and significantly distorts the received signal~\cite{book_synch_imp}. Moreover, due to its time varying nature~\cite{Mehrpouyan}, it is difficult to accurately track and compensate the deteriorating effect of PHN at the receiver.

It is well-known that parameter estimation accuracy can be significantly enhanced if it is carried out jointly with data detection~\cite{Herzet1}. As such, many iterative receiver structures have been proposed that utilize forward error correcting (FEC) codes to perform joint synchronization parameter estimation and data detection. Such iterative receivers were first proposed in~\cite{Lottici} and have since been formalized in~\cite{Noels} with the use of the \emph{expectation-maximization (EM)} framework. In~\cite{Shehata}, a coded iterative structure based on the EM algorithm for tracking PHN in single-input single-output (SISO) systems is proposed. However, the performance of the approach in~\cite{Shehata} degrades with increasing block length and it is also not applicable to MIMO systems. Code-aided synchronization based on the EM framework for joint channel estimation and frequency/time synchronization for MIMO systems is considered in~\cite{HenkSimoens}. However, in~\cite{HenkSimoens}, the synchronization parameters are assumed to be constant and deterministic over the length of a block, which is not a valid assumption for time varying PHN. It is also important to note that unlike SISO systems, MIMO systems may need to employ independent oscillators at each transmit and receive antennas, e.g., for \emph{line-of-sight (LoS)} MIMO systems, where the antennas are positioned far apart from one another~\cite{article_MIMO_LoS,Mehrpouyan}\footnote{For a $4\times 4$ LoS MIMO system operating at $10$ GHz and with a transmitter and receiver distance of $2$ km, the optimal antenna spacing is $3.8$ m~\cite{article_MIMO_LoS}.\vspace{-0pt}} or in the case of multi-user MIMO systems, where independent oscillators are used by different users~\cite{article_PHASE_N_MIMO_OFDM_VIII}. Thus, the signals at the MIMO receiver may be affected by multiple PHN processes that need to be jointly tracked. Although PHN estimation in MIMO systems has been considered in~\cite{Hadaschik,Mehrpouyan, nasir_2013}, these approaches do not address the problem of joint PHN estimation and data detection. Consequently, the performances of the schemes in~\cite{Hadaschik,Mehrpouyan, nasir_2013} are inferior to the scheme proposed here.

\vspace{-10pt}
\subsection{Contributions}

In this paper, the problem of joint iterative coded PHN estimation and data detection in MIMO systems is addressed. The paper's main contributions are summarized as follows:
\begin{itemize}
\item An EM-based receiver for joint PHN estimation and data detection for BICM-MIMO systems is proposed. The EM approach is iteratively applied over the frame, where \emph{low density parity check (LDPC)} codes are used to enhance both data-detection and PHN estimation. To the best of the authors' knowledge, this is the first work that proposes such a receiver structure for tracking nondeterministic parameters, e.g., PHN, over a transmission frame for MIMO systems.
\item Unlike the results in~\cite{Shehata}, it is analytically shown that at high signal-to-noise ratios (SNRs), a maximum a posteriori (MAP) estimator can be used to carry out the maximization step of the EM algorithm. To reduce the computational complexity of the proposed iterative receiver, instead of a MAP estimator, an \emph{extended Kalman filter-smoother (EKFS)} is applied.
\item Extensive simulations are carried out for different PHN variances to show that the performance of a MIMO system employing the proposed receiver structure is very close to the ideal case of perfect synchronization. These simulations also demonstrate that the proposed joint estimation and detection approach is far superior to schemes that perform estimation and detection separately, e.g.,~\cite{nasir_2013}.
    %The \emph{mean square error (MSE)} performance of the proposed scheme for tracking PHN over a frame is also investigated and is shown to improve with every EM iteration.
\end{itemize}

\vspace{-10pt}
\subsection{Organization}
The remainder of the paper is organized as follows: Section \ref{sec:SystemModel} presents the system model while the proposed EM-based PHN estimator is derived in Section \ref{sec:EMalg}. Section \ref{sec:Apps} presents the structure of the iterative detector. Section \ref{sec:complexity} presents the complexity analysis for the proposed receiver structure. Finally, Section \ref{sec:SimRes} presents the results of our extensive simulations.
\vspace{-10pt}
\subsection{Notations}
 Superscripts $(\cdot)^{H}$ and $(\cdot)^{T}$ denote the conjugate transpose and transpose operators, respectively. Bold face small letters, e.g., $\mathbf{x}$, are used for vectors and bold face capital letters, e.g., $\mathbf{X}$, are used for matrices. $\mathbf{I}_{X\times X}$ and $\mathbf{0}_{X \times X}$ denote the $X\times X$ identity and all zero matrices, respectively. $\text{diag}(\mathbf{x})$ denotes a diagonal matrix, where the diagonal elements are given by the vector $\mathbf{x}$. $\|\cdot\|$, \text{tr}($\cdot$), $\mathbb{E}\{\cdot\}$, $\Re\{\cdot\}$, and $\Im\{\cdot\}$ denote the Frobenius norm, trace, expectation, real, and imaginary operators, respectively. $p(x|y)$ denotes the probability distribution function of $x$ given $y$. Finally, $\mathcal{N}\left(\mu, \sigma^2\right)$ and $\mathcal{CN}\left(\mu, \sigma^2\right)$ denote real and complex Gaussian distributions, respectively, with mean $\mu$ and variance $\sigma^2$.

\vspace{-0pt}
\section{System Model\label{sec:SystemModel}}
A MIMO system with $N_t$ transmit antennas and $N_r$ receive antennas is considered. At the transmitter, the coded bits are interleaved and modulated onto an $M$-point quadrature amplitude modulation ($M$-QAM) constellation denoted by $\Omega$. Subsequently, using spatial multiplexing, the symbols are transmitted simultaneously from $N_t$ antennas. Frame based transmission is considered, where $L_f$ denotes the frame length. In this paper, the following set of assumptions is adopted:
\begin{enumerate}[{A}1.]
\item Quasi-static block fading channels are considered.
\item Training sequences transmitted at the beginning of each frame are used to estimate the channel parameters using the algorithm in~\cite{Mehrpouyan}. Thus, the subsequent analysis is based on the assumption that the MIMO channel matrix $\mathbf{H}$ is known. However, in Section~\ref{sec:SimRes}, extensive simulations are carried out by estimating the channel parameters using the algorithm in~\cite{Mehrpouyan}. The assumption of known channel parameters is justified since the topic of joint channel and PHN estimation using a \emph{known training sequence} is addressed in~\cite{Mehrpouyan}.
\item To ensure generality and also applicability of the proposed scheme to LoS and multi-user MIMO systems, it is assumed that independent oscillators are deployed at each transmit and receive antenna.
\item PHN is modeled as a discrete-time Wiener process, i.e., PHN at time $k$, $\theta(k)$ is given by
\begin{align*}
\theta(k)=\theta(k-1)+\Delta(k),
\end{align*}
where $\Delta(k)$ is the PHN innovation~\cite{Demir}.
\end{enumerate}
Assumptions A1 and A3 are in line with previous PHN estimation algorithms in MIMO systems~\cite{Hadaschik,Mehrpouyan, nasir_2013}. Moreover, both assumptions are justifiable in many practical scenarios, e.g., in MIMO microwave backhaul links~\cite{Wells-2010}, where the channel parameters change much more slowly than the PHN process and independent oscillators are used at each antenna due to the large antenna spacing ($2$-$4$ meters).

The received signal vector at time instant $k$, $\mathbf y(k) \triangleq \left[ y_1(k),\allowbreak y_2(k),  \dots, y_{N_r}(k)  \right]^T$, is given by\vspace{-0pt}
\begin{align}\label{ref:obs1}
 \mathbf y(k) =& \boldsymbol \Gamma^{[r]}(k) \mathbf H \boldsymbol \Gamma^{[t]}(k) \mathbf s(k) + \mathbf w(k),
\end{align}
where
\begin{itemize}
\item $\boldsymbol \Gamma^{[r]}(k)\triangleq \textrm{diag}\big(e^{j\theta_1^{[r]}(k)},\dots,e^{j\theta_{N_r}^{[r]}(k)} \big)$ and $\boldsymbol \Gamma^{[t]}(k)\triangleq \textrm{diag}\big(e^{j\theta_1^{[t]}(k)},\dots,e^{j\theta_{N_t}^{[t]}(k)} \big)$ are $N_r\times N_r$ and $N_t\times N_t$ diagonal matrices, respectively,
\item $\theta_\ell^{[r]}(k)$ and $\theta_m^{[t]}(k)$ denote the PHN process at the $\ell$th receive and $m$th transmit antennas, respectively,

\item $\mathbf H \triangleq \left[ \mathbf h_1, \dots, \mathbf h_{N_t}  \right]$ with $\mathbf h_{\ell} \triangleq \left[ h_{\ell,1}, \dots,  h_{\ell,N_t}  \right]^T$ is the $N_r\times N_t$ MIMO channel matrix,

\item $h_{\ell,m}$, for $\ell=1,\dots,N_r$ and $m=1,\dots,N_t$, denotes the channel parameter for the $m$th receive and $\ell$th  transmit antennas pair that is assumed to be distributed as $h_{\ell,m}\sim\mathcal{CN}(\mu_{h_{m,\ell}}, \sigma^2_{h_{m,\ell}})$,
\item $\mathbf s(k) \triangleq\left[ s_1(k),  \dots, s_{N_t}(k)  \right]^T$ is the vector of transmitted symbols,
\item $\mathbf w(k) \triangleq\left[ w_1(k),  \dots, w_{N_r}(k)  \right]^T$ is the vector of the zero-mean additive white Gaussian noise (AWGN) at the receiver, i.e., $w_{m}\sim\mathcal{CN}(0, \sigma^2_{w_m})$.
\end{itemize}

%\vspace{-0pt}
%\begin{align}\label{eq:pn_model}
%\theta_\ell^{[r]}(k)=\theta_\ell^{[r]}(k-1)+\Delta_\ell^{[r]}(k), \qquad
%\theta_m^{[t]}(k)=\theta_m^{[t]}(k-1)+\Delta_m^{[t]}(k),
%\end{align}
%where $\Delta_\ell^{[r]}(k)$ and  $\Delta_m^{[t]}(k)$ are the phase innovations for the $\ell$th receiver and $m$th transmit antenna, respectively, with $\Delta_\ell^{[r]}(k)\thicksim \mathcal N(0,\sigma^2_{\Delta})$ and $\Delta_m^{[t]}(k)\thicksim \mathcal N(0,\sigma^2_{\Delta})$.

\vspace{-0pt}
\section{The Proposed EM-Based PHN Estimator\label{sec:EMalg}}

The EM algorithm consists of the \emph{expectation step (E-step)} and the \emph{maximization step (M-step)} \cite{Herzet1} that for the $i$th EM iteration are given by\vspace{-0pt}
\begin{align}
\label{eq:E-stepMIMO}
\mathrm{Q}\Big( {\boldsymbol \Theta}| \hat{\boldsymbol \Theta}^{(i-1)}\Big) =&
 \mathbb{E}_{\mathbf S|\mathbf Y,\hat{\boldsymbol \Theta}^{(i-1)}}\Big\{\ln p(\mathbf Y|\mathbf S, {\boldsymbol \Theta})\Big\}+\ln p(\boldsymbol \Theta),\\
 \label{eq:M-stepMIMO}
 \hat{\boldsymbol \Theta}^{(i)}=&\arg\max_{ {\boldsymbol \Theta}}\Big\{\mathrm{Q}\Big( {\boldsymbol \Theta}| \hat{\boldsymbol \Theta}^{(i-1)}\Big)\Big\},
\end{align}
respectively. In \eqref{eq:E-stepMIMO} and \eqref{eq:M-stepMIMO}, $\left[\mathbf Y\right]_{N_r \times L_f}\! \triangleq\! [\mathbf y(1) ,  \mathbf y(2) , \dots ,\mathbf y(L_f)]$ and $[\boldsymbol \Theta]_{(Nr+Nt)\times L_f} \triangleq [\boldsymbol \theta(1), \boldsymbol \theta(2),\hdots,\allowbreak\boldsymbol   \theta(L_f)]$. Moreover, since the convergence of the EM algorithm is highly dependent on the initialization process, we propose to transmit pilot symbols every $p_r$ symbol within each transmission frame. In the subsequent subsections, the E- and M-steps of the EM algorithm for coded MIMO systems are derived.

\vspace{-2pt}
\subsection{E-step \label{sec:Estep}}
The MIMO received signal vector $\mathbf y(k)$ in \eqref{ref:obs1} can be rewritten as\vspace{-0pt}
\begin{align}\label{eq:received_signal2}
 \mathbf y(k) =&  \mathbf X(k) \mathbf s(k) + \mathbf w(k),
\end{align}
where $\mathbf X(k) \triangleq   \boldsymbol \Gamma^{[r]}(k) \mathbf H \boldsymbol \Gamma^{[t]}(k)$. Accordingly, using straightforward algebraic manipulations, the log likelihood function (LLF) of the received signal matrix, $\mathbf{Y}$, given the transmitted data, $\mathbf S$, and the PHN process, $ {\boldsymbol \Theta}$, can be determined to be proportional to\vspace{-0pt}
\begin{align}
\ln p(\mathbf Y|\mathbf S, {\boldsymbol \Theta})\varpropto&-\sum_{k=1}^{L_f} \left\|\mathbf y(k)-\mathbf X(k)\mathbf s(k)  \right\|^2\notag\\
\varpropto& 2\Re \bigg\{\sum_{k=1}^{L_f} \textrm{tr}\Big(\mathbf y(k) \mathbf s^H(k) \mathbf X^H(k) \Big)\bigg\}
- \sum_{k=1}^{L_f} \textrm{tr}\Big(\mathbf X(k) \mathbf s(k) \mathbf s^H(k) \mathbf X^H(k) \Big).\label{eqn:channelLF}
\end{align}
Using \eqref{eqn:channelLF}, the E-step in \eqref{eq:E-stepMIMO} can be rewritten as\vspace{-0pt}
\begin{align}
\mathrm{Q}\Big( {\boldsymbol \Theta}|\hat{\boldsymbol \Theta}^{(i-1)}\Big)\propto&2\Re\bigg\{ \sum_{k=1}^{L_f} \textrm{tr}\Big(\mathbf y(k) \boldsymbol \alpha^H(k) \mathbf X^H(k) \Big)\bigg\}
- \sum_{k=1}^{L_f} \textrm{tr}\Big(\mathbf X(k)  \mathbf B(k) \mathbf X^H(k) \Big)
+\ln p(\mathbf {\boldsymbol \Theta}), \label{eqn:Estepfinal}
\end{align}
where
\begin{align}
\boldsymbol \alpha(k)&\triangleq \displaystyle\sum_{\mathbf {a}_n \in  M^{N_t}}\mathbf {a}_n p\left(\mathbf {s}(k)=\mathbf {a}_n|\mathbf Y,\hat{\boldsymbol \Theta}^{(i-1)}\right)\label{eqn:alpha}\\
\mathbf B(k)&\triangleq \displaystyle\sum_{\mathbf {a}_n \in  M^{N_t}}\mathbf {a}_n \mathbf {a}^H_n p\left(\mathbf {s}(k)=\mathbf {a}_n|\mathbf Y,\hat{\boldsymbol \Theta}^{(i-1)}\right).
\end{align}
Here, $\boldsymbol \alpha(k)$ denotes the marginal posterior mean of the coded symbol vector at time $k$, i.e., the soft decisions, and $p\left(\mathbf {s}(k)=\mathbf {a}_n|\mathbf Y,\hat{\boldsymbol \Theta}^{(i-1)}\right)$ denotes the a posteriori probabilities (APPs) of the coded symbol vector given $\mathbf Y$ and $\hat{\boldsymbol \Theta}^{(i-1)}$. Note that at high SNR, i.e., $p\left(\mathbf {s}(k)=\mathbf {a}_n|\mathbf Y,\hat{\boldsymbol \Theta}^{(i-1)}\right)=0$, $\forall k\neq n$, $\boldsymbol \alpha(k)=\mathbf s(k)$ and $\mathbf B(k)=\mathbf {s}(k)\mathbf {s}^H(k)$.

%, and $\mathbf B(k)$ is an ($N_t\times N_t$) matrix.

%The computation of the true posterior probabilities has a complexity that increases exponentially with the frame length $L_f$. Therefore, a near optimal iterative detector, operating according to the turbo principle  \cite{Duman},\cite{Boutros},\cite{HenkSimoens} is used to obtain the marginal a posteriori bit probabilities given the PHN estimates and the channel gain matrix $\mathbf H$. The block diagram of the proposed EM-based receiver structure, including both the EKFS and the iterative detector, is shown in Fig. \ref{fig:fig3}.

\vspace{-2pt}
\subsection{M-step\label{sec:Mstep}}

In this section, we seek to show that a MAP estimator can be applied to carry out the M-step of the proposed EM-based PHN estimator at high SNR. Given the observation matrix $\mathbf Y$, the MAP estimate of $\hat{\boldsymbol \Theta}$ is given by \cite{SMKay}\vspace{-0pt}
\begin{align}\label{eq:e-step}
\hat{ \boldsymbol \Theta}= \textrm{arg}, \max_{ { \boldsymbol{\Theta}}} \big\{\ln p(\mathbf Y| {\boldsymbol{\Theta}}, \mathbf S= \mathbf A)+\ln p(\mathbf S= \mathbf A) + \ln p(\boldsymbol{\Theta})\big\},
\end{align}
where $[\mathbf A]_{N_t\times L_f}\triangleq  \left[\boldsymbol \alpha(1) , \boldsymbol \alpha(2) , \dots ,\boldsymbol \alpha(L_f)\right]$.
%where $p(\mathbf Y| {\boldsymbol \Theta})=\displaystyle\sum_{\mathbf S} p(\mathbf S) p(\mathbf Y|\mathbf S, {\boldsymbol \Theta})$.
%The EM algorithm converges to the MAP solution if the initial estimates are sufficiently close to the true values. Thus,
%As shown in \eqref{eq:M-stepMIMO}, the M-step maximizes the E-step in \eqref{eqn:Estepfinal} with respect to the parameter of interest, $\boldsymbol\Theta$.
%
%
%\begin{align}
%\hat{ \boldsymbol{\Theta}}
%=& \textrm{arg}, \max_{ { \boldsymbol{\Theta}}} \{\ln p(\mathbf Y| {\boldsymbol{\Theta}}, \mathbf S= \mathbf A)
%\notag\\
%&\hspace{+2cm}+\ln p(\mathbf S= \mathbf A) + \ln p(\boldsymbol{\Theta})\}\label{eqn:MAPln2}
%\end{align}
Assuming equally probable transmitted symbols, at high SNR, i.e., $p\left(\mathbf {s}(k)=\mathbf {a}_n|\mathbf Y,\hat{\boldsymbol \Theta}^{(i-1)}\right)=0$, $\forall k\neq n$, \eqref{eq:e-step} can be rewritten as\vspace{-0pt}
\begin{subequations}
\begin{align}
\hat{ \boldsymbol{\Theta}}
 =&\textrm{arg} \max_{ { \boldsymbol{\Theta}}} \{\ln p(\mathbf Y| {\boldsymbol{\Theta}}, \mathbf S= \mathbf A) + \ln p(\boldsymbol{\Theta})\}\label{eqn:MAPln2}\\
=& \textrm{arg} \max_{ { \boldsymbol{\Theta}}} \Bigg\{2\Re\bigg\{ \sum_{k=1}^{L_f} \textrm{tr}\Big(\mathbf y(k) \mathbf {s}^H(k) \mathbf X^H(k) \Big)\bigg\}
- \sum_{k=1}^{L_f} \textrm{tr}\Big(\mathbf X(k) \mathbf {s}(k) \mathbf {s}^H(k) \mathbf X^H(k) \Big)+\ln p(\boldsymbol{\Theta})\Bigg\}
\label{eqn:channelLF2}.
\end{align}
\end{subequations}
%Eq. \eqref{eqn:MAPln2} follows from the assumption of equally probable transmitted symbols and \eqref{eqn:channelLF2} follows from the application of \eqref{eqn:channelLF} and the high SNR assumption. Furthermore, since at high SNRs the soft-decisions, $\boldsymbol \alpha(k)$ reach their true values, $\mathbf{s}(k)$, $\forall k$, it can be concluded that $\boldsymbol \alpha(k)\boldsymbol \alpha^H(k)\approx \mathbf B(k)$.
Based on the discussion following \eqref{eqn:Estepfinal}, the results in \eqref{eqn:Estepfinal} and \eqref{eqn:channelLF2} are equivalent at high SNR. Thus, (\ref{eqn:channelLF2}) is in fact maximizing the E-Step in (\ref{eqn:Estepfinal}), which indicates that a MAP estimator can be applied to carry out the M-step of the proposed EM algorithm at high SNR.

It is important to indicate that although a MAP estimator can be applied to carry out the M-step of the EM algorithm, solving for the MAP solution in \eqref{eqn:channelLF2} requires a multidimensional exhaustive search that is computationally very intensive.
%\footnote{Following the methodology in \cite{nasir_2013}, the computational complexity of the EKFS is $2.2 \times 10^{18}$ less than the MAP estimator.}
Thus, we propose to apply a Kalman filter, which is an optimal linear minimum mean square error estimator \cite{SMKay}, to carry out the maximization step of the EM algorithm and reduce its computational complexity. The set of EKFS equations are provided in following subsection.

%The \emph{observation} equation for the Kalman filter is given by \eqref{eq:received_signal2} while based on the assumption of Weiner PHN, the \emph{state} equation at time $k$ is given by\vspace{-12pt}
%\begin{align}\label{eqn:state3}
%  \boldsymbol{\theta}(k)=&\boldsymbol{\theta}(k-1)+\boldsymbol{\Delta}(k),
%\end{align}
%where $\boldsymbol{\Delta}(k)\triangleq\left[ \Delta_1^{[r]}(k), \dots, \Delta_{N_r}^{[r]}(k), \Delta_1^{[t]}(k), \dots, \Delta_{N_t}^{[t]}(k)\right]^T$, $\Delta_{\ell}^{[r]}$ and $\Delta_{m}^{[t]}$ are the PHN innovations at the transmitter and receiver oscillators, respectively.
%
%
%The transmitted symbols $\mathbf s(k)$, can be replaced by their a posteriori means, i.e., soft-decisions $\boldsymbol \alpha(k)$ computed at the E-step using the iterative detector. Since the \emph{observations} are a nonlinear function of the parameters of interest, $\boldsymbol{\theta}_k$, an \emph{extended} Kalman filter-smoother needs to be used instead to carry out the M-step of the EM algorithm \cite{SMKay}.

%and has a similar structure to the MAP estimator \cite{SMKay}. As a result, it can be applied instead of a MAP estimator to carry out the M-step of the EM algorithm.

\vspace{-0pt}
\subsection{The Extended Kalman Filter-Smoother\label{sec:EKFS}}

In this subsection a low complexity EKFS is applied to carry out the M-step of the EM algorithm. We first note that due to a phase ambiguity, the $N_r+N_t$ PHN parameters $\boldsymbol{\theta}(k)$ cannot be jointly estimated \cite{Hadaschik}. Instead, by arbitrarily selecting the PHN process, $\theta_{N_t}^{[t]}(k)$, as a reference PHN value, $\mathbf X(k)$ in (\ref{eq:received_signal2}) can be rewritten as \cite{nasir_2013}\vspace{-0pt}
\begin{align}\label{eq:new_model}
\mathbf X(k)
=\tilde {\boldsymbol \Gamma}^{[r]}(k) \mathbf H \tilde {\boldsymbol \Gamma}^{[t]}(k),
\end{align}
where
$\tilde{\boldsymbol \Gamma}^{[r]}(k)\triangleq \textrm{diag}\Big\{e^{j(\phi_1(k))}, \dots, e^{j(\phi_{N_r}(k))} \Big\}$,
$\tilde{\boldsymbol \Gamma}^{[t]}(k)\triangleq \textrm{diag}\Big\{e^{j(\phi_{N_r+1}(k))}, \dots, e^{j(\phi_{N_r+N_t-1}(k))}, 1 \Big\}$, $\boldsymbol{\phi}(k) \triangleq \left[ \phi_1(k),\hdots,\phi_{N_r+N_t-1}(k)\right]^T$, and
        \begin{align*}
        \phi_f(k)\triangleq
        \begin{cases}
            \theta^{[r]}_f+\theta^{[t]}_{N_t}&f=1,\hdots,N_r,\\
            \theta^{[t]}_{f-N_r}-\theta^{[t]}_{N_t}&f=N_r+1,\hdots,N_r+N_t-1.
        \end{cases}
       \end{align*}
The application of the equivalent signal model in \eqref{eq:new_model} instead of \eqref{ref:obs1} eliminates the ambiguity associated with the estimation of PHN parameters. Subsequently, the state and observation equations can be determined as
\begin{align}
\boldsymbol{\phi}(k)=&\boldsymbol{\phi}(k-1)+\tilde{\boldsymbol{\Delta}}(k)\label{eqn:state4},\\
\mathbf y(k) \approx& \mathbf z(\boldsymbol{\phi}(k))+ \mathbf w(k)\label{eqn:obs4},
\end{align}
where $\mathbf z(\boldsymbol{\phi}(k))\triangleq\left[ z_1(\boldsymbol{\phi}(k)), z_2(\boldsymbol{\phi}(k)), \dots,  z_{N_r}(\boldsymbol{\phi}(k)) \right]^T=\tilde {\boldsymbol \Gamma}^{[r]}(k) \mathbf H \tilde {\boldsymbol \Gamma}^{[t]}(k) \boldsymbol \alpha(k)$, $\tilde{\boldsymbol{\Delta}}(k)\triangleq\big[\Delta_1^{[r]}(k)+\Delta_{N_t}^{[t]}(k), \dots, $ $ \Delta_{N_r}^{[r]}(k)+\Delta_{N_t}^{[t]}(k), \Delta_1^{[t]}(k)-\Delta_{N_t}^{[t]}(k), \dots, \Delta_{N_t-1}^{[t]}(k)-\Delta_{N_t}^{[t]}(k)\big]^T$, and $\Delta_\ell^{[r]}(k)$ and $\Delta_m^{[t]}(k)$ are the PHN innovations corresponding to the $\ell$th receive and $m$th transmit antenna PHN parameters, $\theta_\ell^{[r]}(k)$ and $\theta_m^{[t]}(k)$, respectively \cite{article_PHASE_N_MODEL_XII}. $\Delta_\ell^{[r]}(k)$ and $\Delta_m^{[t]}(k)$, $\forall \ell,m$, are modelled as real Gaussian processes, i.e., $\Delta_\ell^{[r]}(k),\Delta_m^{[t]}(k)\sim\mathcal{N}\left(0, \sigma_\Delta^2\right)$, $\forall \ell,m$ \cite{article_PHASE_N_MODEL_XII, Mehrpouyan}. Moreover, $\boldsymbol \alpha(k)$ is defined after \eqref{eqn:Estepfinal}. Note that since $z_m(\boldsymbol{\phi}(k))$ is a nonlinear function of $\boldsymbol{\phi}(k)$ the extended Kalman filter is applied here.

Recall that \emph{complex} channel parameters (consisting of the channel gain and \emph{phase}) are estimated at the start of the frame via a training sequence and the algorithm in \cite{Mehrpouyan}. Consequently, the EKFS is initialized with the state estimate $\hat {\boldsymbol \phi}(0)=\mathbf{0}_{(N_t+N_r-1)\times 1}$ and the error covariance matrix $\hat {\mathbf M}(0)=2\sigma^2_\Delta\mathbf I$. Subsequently, the EKFS determines the {a priori} state vector, $\hat {\boldsymbol \phi}^-(k)$, and the {a priori} error covariance matrix, $\left[\hat {\mathbf M}^{^-}(k)\right]_{(N_r+N_t-1)\times (N_r+N_t-1)}$, using $\hat {\boldsymbol \phi}^-(k)=\hat {\boldsymbol \phi}(k-1)$ and $\hat {\mathbf M}^{^-}(k) = \hat {\mathbf M}(k)+ 2 \sigma^2_\Delta \mathbf I$, respectively. The EKFS linearizes the nonlinear function $\mathbf z(\boldsymbol{\phi}(k))$ in (\ref{eqn:obs4}) about the a priori estimate of the state vector via\vspace{-0pt}
\begin{align}
\mathbf z( {\boldsymbol \phi}(k))\approx& \mathbf z(\hat {\boldsymbol \phi}^-(k)) + \dot{\mathbf Z}(k)(\boldsymbol \phi(k)-\hat {\boldsymbol \phi}^-(k)),\label{eqn:phiTaylor}
\end{align}
where the Jacobian matrix with respect to $\boldsymbol \phi$ at time instance $k$, $\left[\dot{\mathbf Z}(k)\right]_{N_r\times(N_r+N_t-1)}$, can be determined as\vspace{-12pt}
\begin{align}\label{eq:jacobian}
\dot{\mathbf Z}(k)\triangleq&  \frac{\partial {\mathbf z}}{\partial \boldsymbol\phi(k)}\Big |_{\hat {\boldsymbol \phi}^-(k)}= \left[\dot{\mathbf Z}_1(k), \dot{\mathbf Z}_2(k)\right].
\end{align}
In \eqref{eq:jacobian}, the matrix $\left[\dot{\mathbf Z}_1(k)\right]_{N_r\times N_r}$ is given by
\begin{align}
 \dot{\mathbf Z}_1(k) \triangleq& \textrm{diag}\Big\{j z_1(\hat {\boldsymbol \phi}^-(k)),\dots,j z_{N_r}(\hat {\boldsymbol \phi}^-(k))\Big\},
\end{align}
and the matrix $\left[\dot{\mathbf Z}_2(k)\right]_{N_r\times(N_t-1)}$ is determined as
\begin{align}
\label{eqn_dbl_x}
\dot{\mathbf Z}_2(k) \!\triangleq\!\left[\!\! \begin{array}{ccc} \!\!
  jh_{11}e^{j\hat {\phi}^{^-}_1(k)}\alpha_1(k)e^{j\hat {\phi}^{^-}_{N_r+1}(k)} \!&\!\! \dots \!\!&\! jh_{1(N_t-1)}e^{j\hat {\phi}^{^-}_{1}(k)}\alpha_{N_t-1}(k)e^{j\hat {\phi}^{^-}_{N_r+N_t-1}(k)} \\
 \!\!jh_{21}e^{j\hat {\phi}^{^-}_2(k)}\alpha_1(k)e^{j\hat {\phi}^{^-}_{N_r+1}(k)} \!&\!\! \dots \!\!&\! jh_{2(N_t-1)}e^{j\hat {\phi}^{^-}_{2}(k)}\alpha_{N_t-1}(k)e^{j\hat {\phi}^{^-}_{N_r+N_t-1}(k)} \\
\!\!\vdots  \!&\!\! \ddots \!\!&\! \vdots  \\
   \!\! jh_{N_r1}e^{j\hat {\phi}^{^-}_{N_r}(k)}\alpha_1(k)e^{j\hat {\phi}^{^-}_{N_r+1}(k)} \!&\!\! \dots \ \!\!&\! jh_{N_r(N_t-1)}e^{j\hat {\phi}^{^-}_{N_r}(k)}\alpha_{N_t-1}(k)e^{j\hat {\phi}^{^-}_{N_r+N_t-1}(k)}
\end{array} \!\!\right].
\end{align}

After the observation, the posteriori estimate of the state vector, denoted by $\hat {\boldsymbol \phi}^+(k)$, and the posteriori error covariance matrix, denoted by $\hat {\mathbf M}^{^+}(k)$, are determined as
\begin{subequations}
\begin{align}
\hat {\boldsymbol \phi}^+(k)=&\hat {\boldsymbol \phi}^-(k) + \Re \{\mathbf K(k)(\mathbf y(k)-\mathbf z(\hat{\boldsymbol \phi}^-(k))) \}\label{eqn:ekfs-state},\\
\hat {\mathbf M}^{^+}(k) =& \Big(\mathbf I-\Re \{\mathbf K(k) \dot{\mathbf Z}(k)\}\Big)\hat {\mathbf M}^{^-}(k),\label{eqn:ekfs-cov}
\end{align}
\end{subequations}
where $\mathbf K(k)= \hat {\mathbf M}^{^-}(k)\dot{\mathbf Z}(k)^H\Big(\mathbf C_w+\dot{\mathbf Z}(k) \hat {\mathbf M}^{^-}(k)\dot{\mathbf Z}(k)^H \Big)^{-1}$ is the $(N_r+N_t-1)\times N_r$ Kalman Gain matrix and $\mathbf C_w=\Big(\frac{\sigma^2_w}{2} + j\frac{\sigma^2_w}{2}\Big) \mathbf I$ is the observation noise covariance matrix. The EKFS also runs a backward recursion to smooth the a posteriori estimates of the state statistics over the block. The smoothed estimate of the PHN vector, $\hat {\boldsymbol \phi}(k)$, and the error covariance matrix $\hat {\mathbf M}(k)$ are given by\vspace{-0pt}
\begin{subequations}
\begin{align}
\!\!\hat {\boldsymbol \phi}(k)=&\hat {\boldsymbol \phi}^+(k) \!+ \hat {\mathbf M}^{^+}(k) (\hat {\mathbf M}^{^-}(k+1))^{-1}
\Big(\hat {\boldsymbol \phi}(k+1)\!-\!\hat {\boldsymbol \phi}^-(k+1)\Big) \label{eqn:eks-state-rev}\\
\!\!\hat {\mathbf M}(k)=&\hat {\mathbf M}^{^+}(k)\!+\hat {\mathbf M}^{^+}(k) (\hat {\mathbf M}^{^-}(k+1))^{-1}
\Big(\hat {\mathbf M}(k+1)\!-\hat {\mathbf M}^{^-}(k+1)\Big)
\Big(\hat {\mathbf M}^{^+}(k) (\hat {\mathbf M}^{^-}(k+1))^{-1}\Big)^T.\label{eqn:eks-cov-rev}
\end{align}
\end{subequations}
After the backward recursion is completed, the block of PHN estimates, $\hat {\boldsymbol \phi}(k)$, for $k=1,2,\dots,L_f$, are fed to the iterative detector, which is presented in the following section.

 %for the next EM algorithm iteration as shown in Fig. \ref{fig:fig3}.

\begin{figure}[t]%t h b
\begin{center}
\vspace{-0.0cm}
\includegraphics[width=18cm]{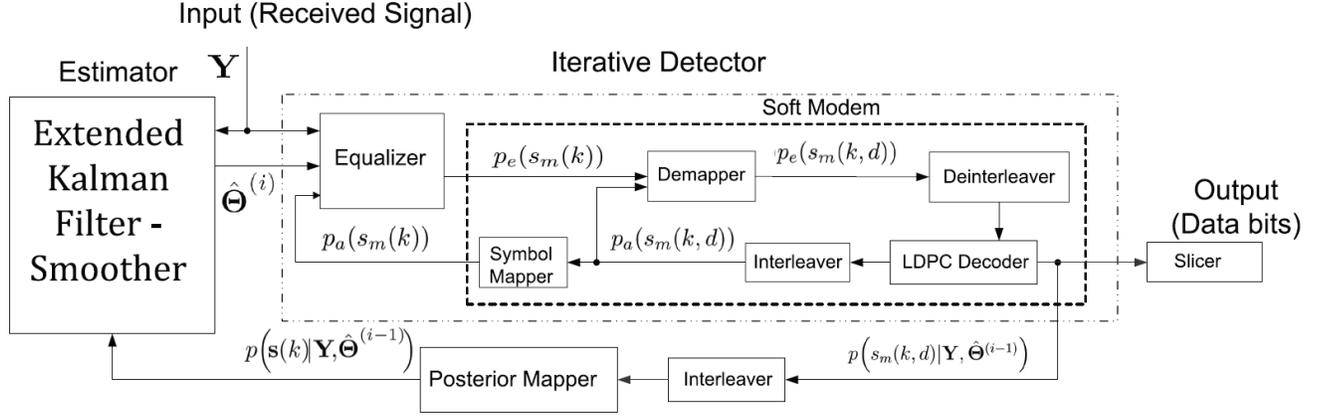}
\caption{Block diagram of the receiver structure, where $d$ denotes the number of iterations within the soft-modem.}\label{fig:fig3}
\end{center}
%\vspace{-.5cm}
\end{figure}

%\begin{figure}[t]%t h b
%\begin{center}
%\includegraphics[width=14cm]{ReceiverStructureMIMO.pdf}
%%\vspace{-5.3cm}
%\caption{Block diagram of the receiver structure.}\label{fig:fig3}
%\end{center}
%\vspace{-.3cm}
%\end{figure}

\section{Iterative Detector\label{sec:Apps}}
As indicated in Sec. \ref{sec:Mstep}, the soft decisions, $\mathbf A$, are required by the EKFS to estimate the PHN parameters. However, the computation of the true posterior probabilities has a complexity that increases exponentially with the frame length $L_f$. Therefore, here, a near optimal iterative detector is used in combination with a soft modulator to map the a posteriori bit probabilities to symbol probabilities and construct the soft decisions \cite{Duman},\cite{Boutros},\cite{HenkSimoens}. The block diagram of the proposed EM-based receiver structure, including both the EKFS and the iterative detector, is shown in Fig.~\ref{fig:fig3}. Note that since an interleaver at the transmitter side is used, the symbols transmitted from each antenna and the bits within each symbol are independent.
%Note that all the blocks in Fig. \ref{fig:fig3} computation is performed over the frame, i.e., $k=1,2,\dots,L_f$.
%It is seen that the iterative detector has 3 nested iterations .
%The first iteration is between the iterative detector (E-step) and  the estimator (M-step). $L_{EM}$ denotes the number of EM algorithm iterations.

In order to describe the structure of the proposed iterative detector, we introduce the following notations:
\begin{itemize}
\item $p(\cdot), p_a(\cdot),$ and $p_e(\cdot)$ denote \emph{a posteriori}, \emph{a priori}, and \emph{extrinsic} probabilities, respectively.
\item $p\Big(s_m(k)|\mathbf Y, \hat{\boldsymbol \Theta}^{(i-1)}\Big)$ is the conditional \emph{a posteriori} probability of the transmitted symbol at the transmit antenna $m$, at time instance $k$, $s_m(k)$. According to the turbo principle, it can be factored as
    \begin{eqnarray}
    p\Big(s_m(k)|\mathbf Y, \hat{\boldsymbol \Theta}^{(i-1)}\Big)&=&C^{(1)}p_a(s_m(k))p_e(s_m(k))\label{eqn:symbPostFact}.
    \end{eqnarray}
    where $p_a(s_m(k))$ is the a priori symbol probability, $p_e(s_m(k))$ is the extrinsic symbol probability, and $C^{(1)}$ is a normalization constant. The conditional terms in the extrinsic and the a priori probabilities are omitted for notational simplicity. Note that, these posterior probabilities are used to construct the soft decision, $\boldsymbol \alpha(k)$,  which is required by the M-step.
\item Similarly, $p\Big(\!s_m(k,d)|\mathbf Y\!,\hat{\boldsymbol \Theta}^{(i-1)}\!\Big)$ denotes the \emph{bit posterior probabilities} of the $d$th bit of the bit sequence mapped to the symbol $s_m(k)$, denoted as $s_m(k,d)$. $p\Big(\!s_m(k,d)|\mathbf Y\!,\hat{\boldsymbol \Theta}^{(i-1)}\!\Big)$ is given as a product of the a priori bit probabilities, $p_a(s_m(k,d))$, the extrinsic bit probabilities $p_e(s_m(k,d))$, and a normalization constant, $C^{(2)}$. Note that the posterior bit probabilities, $p\Big(s_m(k,d)|\mathbf Y,\hat{\boldsymbol \Theta}^{(i-1)}\Big)$, are used for hard decision at the end of a number of EM iterations.
 \item $p\Big(\mathbf Y|\mathbf s(k),\hat{\boldsymbol \Theta}^{(i-1)}\Big)$ denotes the conditional likelihoods of the symbol vectors computed by the equalizer from  the received signal before the execution of the detector iterations.
\end{itemize}

\IncMargin{1em}
\begin{algorithm}[t]
\caption{The Iterative Detector Algorithm for the E-step at the $i$-th EM iteration}
\label{Myalgo1}
\DontPrintSemicolon
\SetKwInOut{Input}{input}
\SetKwInOut{Output}{output}
\BlankLine
\Input{$\hat{\boldsymbol \Theta}^{(i-1)},\mathbf H$}
\Output{$\boldsymbol \alpha(k), \hat{s}_{m}(k,d)$, \qquad $\forall k,m,d$}
$p\Big(\mathbf Y|\mathbf s(k),\hat{\boldsymbol \Theta}^{(i-1)}\Big) \leftarrow (\ref{eqn:channelProbs}),\qquad \forall k$\;
\For{$l :=1$ \KwTo $L_{eq-sm}$  }{
     $p_e(s_m(k)) \leftarrow (\ref{eqn:equalizer}),\qquad \forall k,m$ \;
    \For{$b :=1$ \KwTo $L_{dm-dc}$ }{
     $p_e(s_m(k,d) \leftarrow (\ref{eqn:demapper}),\qquad \forall k,m,d$ \;
     Deinterleave \;
    $ p_a(s_{m}(k,d)) \leftarrow$ LDPC decoder \;
    $ p\Big(s_m(k,d)|\mathbf Y,\hat{\boldsymbol \Theta}^{(i-1)}\Big) \leftarrow$ LDPC decoder \;
    Interleave \;
    }
    $p_a(s_m(k)) \leftarrow (\ref{eqn:symbolmapper}),\qquad \forall k,m$ \;
}
$p\Big(\mathbf s(k)|\mathbf Y, \hat{\boldsymbol \Theta}^{(i-1)}\Big) \leftarrow (\ref{eqn:postmapper}),\qquad \forall k$\;
$\boldsymbol \alpha(k) \leftarrow  p\Big(\mathbf s(k)|\mathbf Y, \hat{\boldsymbol \Theta}^{(i-1)}\Big) ,\qquad \forall k$\;
$\hat{s}_{m}(k,d) \leftarrow p\Big(s_m(k,d)|\mathbf Y,\hat{\boldsymbol \Theta}^{(i-1)}\Big), \qquad \forall k$\;
\end{algorithm}
\DecMargin{1em}

%The iterative detector first computes $M^{N_t}$ conditional likelihoods of the symbol vectors given the PHN estimates by the equalizer block in Fig.~\ref{fig:fig3} prior to the execution of the detector iterations. The equalizer also computes the extrinsic symbol probabilities.

The proposed iterative receiver computes the symbol and bit probabilities according to in Algorithm \ref{Myalgo1} on the next page and by utilizing the following set of equations:\\
\emph{Equalizer:}
\begin{align}
p\Big(\mathbf Y|\mathbf s(k),\hat{\boldsymbol \Theta}^{(i-1)}\Big)&=p\Big(\mathbf y(k)|\mathbf s(k),\hat{\boldsymbol \Theta}^{(i-1)}\Big)\nonumber\\
%&=&\frac{1}{(2\pi)^{N_r}\sigma_w^{2N_r}}\exp \Big(-\frac{1}{2\sigma^2_w}|\mathbf y(k)-\mathbf X(k)\mathbf s(k)|^2 \Big).
&=C^{(3)} \exp \Big(-\frac{1}{2\sigma^2_w}|\mathbf y(k)-\mathbf X(k)\mathbf s(k)|^2 \Big),\label{eqn:channelProbs}\\
p_e(s_m(k)=a_n)&=p\Big(\mathbf Y|s_m(k)=a_n,\hat{\boldsymbol \Theta}^{(i-1)}\Big) \nonumber\\
     &= \sum_{\mathbf s(k):s_m(k)=a_n} \Big\{p\Big(\mathbf Y|\mathbf s(k),\hat{\boldsymbol \Theta}^{(i-1)}\Big)\prod_{m'\neq m} p_a(s_{m'}(k))\Big\},\label{eqn:equalizer}
 \end{align}
\emph{Demapper:}
 \begin{align}
p_e(s_m(k,d)=\beta)&=p\Big(\mathbf Y|s_m(k,d)=\beta,\hat{\boldsymbol \Theta}^{(i-1)}\Big) \nonumber\\
     &= \sum_{a_n \in \Omega:a_n(d)=\beta} \Big\{p_e(s_m(k)=a_n)  \prod_{d'\neq d} p_a(s_m(k,d'))\Big\},\label{eqn:demapper}
     \end{align}
\emph{Symbol mapper:}
 \begin{align}
p_a(s_m(k))&=\prod_{d} p_a(s_{m}(k,d)),\label{eqn:symbolmapper}
\end{align}
\emph{Posterior mapper:}
 \begin{align}
p\Big(\mathbf s(k)|\mathbf Y, \hat{\boldsymbol \Theta}^{(i-1)}\Big)&=C^{(4)}p_a(\mathbf s(k))p_e(\mathbf s(k)) \\
&=C^{(5)}p\Big(\mathbf Y|\mathbf s(k),\hat{\boldsymbol \Theta}^{(i-1)}\Big)\prod_{m,d} p_a(s_m(k,d))\label{eqn:postmapper}
\end{align}
where $a_n\in \Omega, n=1,2,\dots,M$, $\beta \in \{0,1\}$, and $C^{(3)}, C^{(4)}$, and $C^{(5)}$ are normalization constants. In Algorithm \ref{Myalgo1}, $L_{eq-sm}$ is the number of iterations between the \emph{equalizer} and the \emph{soft modem} and $L_{dm-dc}$ is the number of iterations between the \emph{demapper} and \emph{LDPC decoder}.

%The a priori symbol probabilities $p_a(s_m(k))$ and bit probabilities $p_a(s_m(k,d))$ are initialized with a uniform distribution at the first EM algorithm iteration. A sufficient number of iterations  inside of the detector is required  for convergence if the detector is reinitialized with uniform probabilities at each EM iteration. Instead, the detector can be initialized with the a priori probabilities obtained at the previous EM iteration.

The a priori symbol probabilities, $p_a(s_m(k))$, and bit probabilities, $p_a(s_m(k,d))$, are initialized with a uniform distribution at the first EM algorithm iteration. Subsequently, the detector is initialized with the a priori probabilities obtained at the previous EM iteration.

\section{Complexity Analysis}\label{sec:complexity}
In this paper, computational complexity is defined as the number of complex additions plus multiplications required to
obtain the PHN estimates at the $i$-th iteration, $\hat{\boldsymbol \Theta}^{(i)}$. Throughout this subsection, the superscripts $(\cdot)^{[M]}$
and $(\cdot)^{[A]}$ are used to denote the number of multiplications
and additions required by each algorithm, respectively. In order to reduce the computational complexity of the MAP estimator, it is assumed that alternating projection is applied to carry out the multidimensional exhaustive search in \eqref{eqn:channelLF2}~\cite{Ziskind-A-88}. Subsequently, the complexity of the MAP estimator in (\ref{eqn:channelLF2}), denoted by $C_{MAP}\triangleq C^{[M]}_{MAP}+C^{[A]}_{MAP}$, can be determined as
\begin{align}
C^{[M]}_{MAP}=&\mathcal N (N_r+N_t)L_f\frac{2\pi}{\kappa}\Big\{1+L_f(\underbrace{N_rN_t+N_r^2N_t}_{\textrm{first factor in } (\ref{eqn:channelLF2})}
+\underbrace{2N_rN_t+N_r^2N_t}_{\textrm{second factor in } (\ref{eqn:channelLF2})}
+\underbrace{N_r^2N_t+N_rN_t^2}_{\mathbf X(k) \textrm{ in } (\ref{eqn:channelLF2})}+C^{[M]}_{\boldsymbol \alpha(k)}) \Big\},\\
C^{[A]}_{MAP}=&\mathcal N (N_r+N_t)L_f\frac{2\pi}{\kappa}\Big\{2+L_f(\underbrace{N_r^2(N_t-1)+N_r}_{\textrm{first factor in } (\ref{eqn:channelLF2})}
+\underbrace{N_r^2(N_t-1)+N_rN_t}_{\textrm{second factor in } (\ref{eqn:channelLF2})}\nonumber\\
&+\underbrace{N_rN_t(N_r+N_t-2)}_{\mathbf X(k) \textrm{ in }(\ref{eqn:channelLF2})}
+C^{[A]}_{\boldsymbol \alpha(k)}) \Big\},
\end{align}
where
\begin{itemize}
\item 
\begin{align*}
C^{[M]}_{\boldsymbol \alpha(k)}=&\underbrace{N_tM^{N_t}}_{(\ref{eqn:alpha})}+M^{N_t}\Big\{\underbrace{N_rN_t+N_r+3}_{(\ref{eqn:channelProbs})}+
\underbrace{N_t\log_2M+2}_{(\ref{eqn:postmapper})}
+L_{eq-sm}\Big[\underbrace{N_tM^{N_t-1}}_{(\ref{eqn:equalizer})}
+\underbrace{N_t}_{(\ref{eqn:symbolmapper})}\nonumber\\
&+L_{dm-dc}\Big(\underbrace{\frac{M}{2}\log_2M}_{(\ref{eqn:demapper})}+\underbrace{L_{dec}N_{var}}_\textrm{LDPC decoder}\Big)\Big]\Big\},
\end{align*}
\item
\begin{align*}
C^{[A]}_{\boldsymbol \alpha(k)}=&\underbrace{N_t(M^{N_t}-1)}_{(\ref{eqn:alpha})}+M^{N_t}\Big\{\underbrace{N_rN_t+N_r-1}_{(\ref{eqn:channelProbs})}
+L_{eq-sm}\Big[\underbrace{M^{N_t-1}-1}_{(\ref{eqn:equalizer})}\nonumber\\
&+L_{dm-dc}\Big(\underbrace{\frac{M}{2}-1}_{(\ref{eqn:demapper})}
+\underbrace{L_{dec}(2N_{check}-1)}_\textrm{LDPC decoder}\Big)\Big]\Big\},
\end{align*}
\item $\mathcal N$ denotes the number of alternating projection cycles used, 
\item $\kappa$ denotes the step size used for the exhaustive search, 
\item $L_{eq-sm}$ is the number of iterations between the equalizer and the soft modem, 
\item $L_{dm-dc}$ is the number of iterations between the demapper and LDPC decoder, and 
\item $N_{var}$ and $N_{check}$ denote the number of variable nodes and check nodes of the regular LDPC code, respectively.
\end{itemize}

The complexity of the proposed EKFS, $C_{EKFS}=C^{[M]}_{EKFS}+C^{[A]}_{EKFS}$, can be calculated as
\begin{align}
C^{[M]}_{EKFS}=&L_f\Big\{\underbrace{2N^2N_r+2N_r^2N+N_r^3}_{\mathbf K(k) \textrm{ below } (\ref{eqn:ekfs-cov})}
+\underbrace{N_r+5N_r(N_t-1)}_{\dot{\mathbf Z}(k) \textrm{ in } (\ref{eq:jacobian})}
+\underbrace{N(N_r+1)}_{(\ref{eqn:ekfs-state})}
+\underbrace{N(NN_r+N^2+1)}_{(\ref{eqn:ekfs-cov})}\nonumber\\
&+\underbrace{N_r^2N_t+N_rN_t^2+N_rN_t}_{\mathbf z(\hat{\boldsymbol \phi}^-(k)) \textrm{ in }(\ref{eqn:ekfs-state})}
+C^{[M]}_{\boldsymbol \alpha(k)}
+\underbrace{N^2+N^3}_{(\ref{eqn:eks-state-rev})}
+\underbrace{2N^3}_{(\ref{eqn:eks-cov-rev})}\Big\},\\
C^{[A]}_{EKFS}=&L_f\Big\{\underbrace{N}_{(\ref{eqn:state4})}
+\underbrace{NN_r(2N+N_r-3)+N_r^2N+N_r^3}_{\mathbf K(k) \textrm{ below }(\ref{eqn:ekfs-cov})}
+\underbrace{N_r(N+1)}_{(\ref{eqn:ekfs-state})}
+\underbrace{N^2(N+N_r-1)}_{(\ref{eqn:ekfs-cov})}\nonumber\\
&+\underbrace{N_rN_t(N_r+N_t-1)-N_r}_{\mathbf z(\hat{\boldsymbol \phi}^-(k)) \textrm{ in }(\ref{eqn:ekfs-state})}
+C^{[A]}_{\boldsymbol \alpha(k)}
+\underbrace{N(N^2+1)}_{(\ref{eqn:eks-state-rev})}
+\underbrace{N^2(2N+1)}_{(\ref{eqn:eks-cov-rev})} \Big\},
\end{align}
where $N\triangleq N_r+N_t-1$.

\begin{table}[t]
\caption {Computational Complexity of the MAP and EKFS Estimators for different number of antennas.} \label{tab:complexity}
\begin{center}
\begin{tabular}{|c|c|c|}
  \hline
  % after \\: \hline or \cline{col1-col2} \cline{col3-col4} ...
  MIMO  & $C_{MAP}$ & $C_{EKFS}$ \\ \hline
  $2\times2$  & 2.81e17 & 3.44e8 \\ \hline
  $4\times4$ & 7.36e21 & 4.47e12 \\ \hline
  $8\times8$ & 6.20e31 & 1.89e22 \\
  \hline
\end{tabular}
\end{center}
\end{table}

\newtheorem{remark}{Remark}
\begin{remark}\label{rem:1}
In Table \ref{tab:complexity}, the computational complexity of the MAP and EKFS estimators are compared against one another for $2\times2$, $4\times4$, and $8\times8$ MIMO systems. To ensure accurate estimation via the MAP estimator, we set $\kappa =10^{-3}$ and $\mathcal N=4$. It is also assumed that a rate $R=7/8$ regular LDPC code \cite{Goddard} with a variable node degree of $N_{var}=4$, check node degree of $N_{check}=32$, and a frame length of $L_f=8176$ is used. In addition, $L_{eq-dm}=L_{dm-dc}=L_{dec}=1$. This approach requires more EM iterations but less detector iterations to converge, which further reduces the computational complexity of the proposed iterative detector significantly. Table \ref{tab:complexity} shows that the proposed soft-input EKFS is significantly less computationally complex than the MAP estimator. For example, for a $4\times4$ MIMO system, the EKFS estimator is $1.6 \times 10^{9}$ times less complex than the MAP estimator.
\end{remark}

\vspace{-0pt}
\section{Simulation Results\label{sec:SimRes}}

%Note that only 1 iteration is allowed inside of the iterative detector for each nested iteration. This alternative approach requires more EM iterations but less detector iterations to converge. In this way, the computational complexity of both the detector and the receiver can be reduced significantly.

In this section, the proposed iterative coded receiver structure is extensively simulated. At the transmitter, data bits are first encoded by a rate $R=7/8$ regular LDPC encoder with variable node degrees of $4$ and check node degree of $32$~\cite{Goddard}. Gray mapping is applied. The number of data bits in each frame, $L_b$ is set to $7154$. In order to enhance bandwidth efficiency, $16$-QAM modulation is employed. Performance is measured as a function of $E_b/N_0$, where $E_b$ denotes the transmitted power per bit and $N_0$ is the AWGN power, i.e., $\sigma^2_w=N_0$. Unless otherwise specified, the number of iterations within the LDPC decoder, $L_{dec}$, is set to $1$. A $2\times2$ MIMO system is used for all simulations and Rician Fading channels are considered, i.e., $\mu_{h_{m,\ell}}=0$, $\sigma^2_{h_{m,\ell}}=1$, $\forall m,\ell$. The MIMO channel matrix is generated as a sum of LoS and \emph{non-line-of-sight (NLoS)} components, where the Rician factor, $\rho$, is set to $2$ dB throughout this section~\cite{article_MIMO_LoS}. The channel parameters are estimated using the approach in~\cite{Mehrpouyan}. Finally, in the initial step, the EKFS is applied to estimate the PHN parameters corresponding to every $p_r$ spaced pilot. Afterwards, using linear interpolation, the PHN corresponding to the remaining symbols are estimated. These PHN values are then used by the proposed iterative detector to initialize the proposed EM-based algorithm.

In order to thoroughly investigate the performance of the proposed receiver, the following specific simulation scenarios are considered:
\begin{enumerate}[\emph{Scenario} 1.]
\item The performance of a MIMO receiver that does not track the PHN parameters is simulated. This scenario is denoted by ``no PHN tracking".
\item The performance of the proposed iterative receiver is compared to that of~\cite{Mehrpouyan}. To ensure a fair comparison, the performance of the algorithm in~\cite{Mehrpouyan} is complemented by FEC, i.e. the above LPDC code is used.\footnote{The results in~\cite{Mehrpouyan} are presented for uncoded MIMO systems.} This scenario is denoted by ``\cite{Mehrpouyan} + FEC".
\item To demonstrate the advantage of the proposed joint PHN estimation and data detection algorithm, the performance of the proposed MIMO receiver is compared to the scenario where PHN estimation and data detection are carried out separately, e.g., the approach in~\cite{nasir_2013}. This scenario is denoted by ``\cite{nasir_2013} disjoint".
%\item PHN estimation and data detection are carried out jointly using the proposed iterative receiver (denoted by ``Proposed receiver"); and
\item As a benchmark the performance of a MIMO system that is affected by no PHN is also presented. This scenario is denoted by ``No PHN".
\end{enumerate}
Since the algorithms in~\cite{Mehrpouyan} and~\cite{nasir_2013} are shown to be superior to the approach in~\cite{Hadaschik}, we do not present any comparison results with respect to~\cite{Hadaschik}.
%It is important to note that the third scenario provides a comparison against the proposed iterative receiver in this paper and the approaches that perform PHN estimation and detection separately, specifically,~\cite{nasir_2013}.

\begin{figure}[t]%t h b
\begin{center}
%\vspace{-4cm}
\includegraphics[width=12cm]{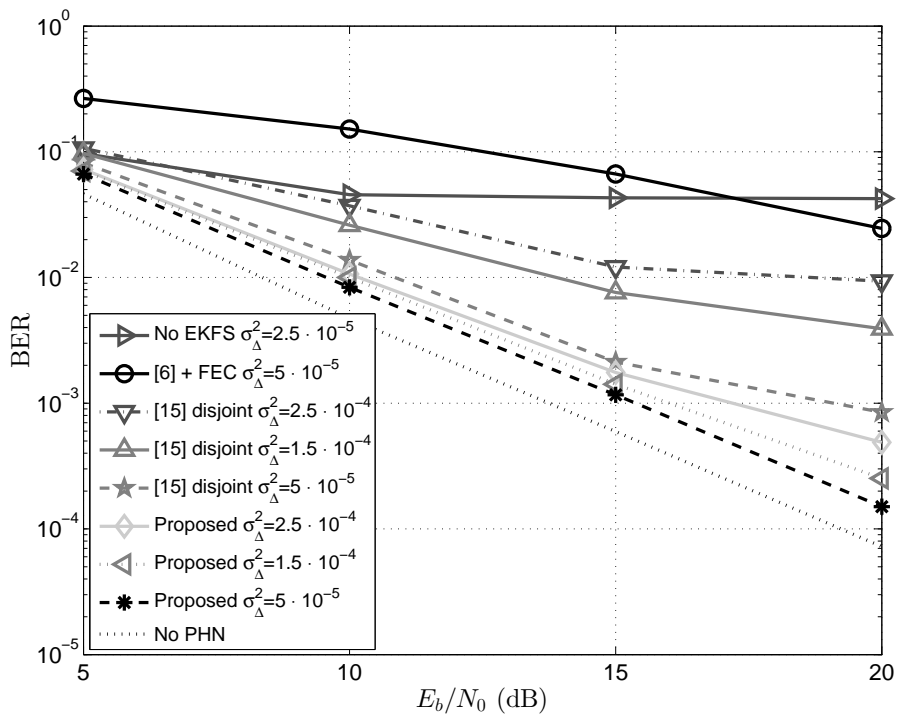}\center
%\vspace{+3.7cm}
\caption{BER performance of the proposed EM-based algorithm ($3$ EM iterations with $L_{dec}=1$).}\label{fig:BERpr14}
\end{center}
\vspace{-.9cm}
\end{figure}

In Fig.~\ref{fig:BERpr14}, the bit error-rate (BER) performance of the proposed EM-receiver is investigated. In this setup, pilot symbols are transmitted every $14$ symbols and only $3$-EM iterations are applied at the receiver. Compared to the ``no PHN" scenario, the proposed EM-based algorithm gives rise to a BER degradation of about $1.5$ dB and $2$ dB for PHN variances of $\sigma^2_\Delta=5\cdot10^{-5}$ rad$^2$ and $\sigma^2_\Delta=1.5\cdot10^{-4}$ rad$^2$, respectively. We also observe that for large PHN variances, i.e., $2.5\cdot 10^{-4}$ rad$^2$, the overall systems suffers from an error floor.\footnote{As shown in \cite{article_PHASE_N_MODEL_XII}, in practice, the PHN innovation variance is small, e.g., for a typical free-running oscillator operating at $2.8$ GHz, the PHN variance is calculated to be $\sigma^2_{\Delta}=5\times10^{-5}$ rad$^2$.} This follows from the fact that the prediction error of the EKFS is determined by the PHN innovation variance. It is also observed that the performance of a MIMO system is significantly degraded when no PHN tracking is applied. More importantly, the results in Fig.~\ref{fig:BERpr14} illustrate that by jointly carrying out PHN estimation and data detection, the proposed receiver results in significant performance gains compared to scenario where PHN estimation and data detection are carried out separately, e.g.,~\cite{nasir_2013}. In fact, on average, a performance gain of more than $10$ dB is observed by applying the proposed receiver.

%However, joint channel and PHN estimation with a FEC decoder is not sufficient to perform close to no PHN case. It states that using joint estimation and detection techniques on top of joint channel and PHN estimation is obligatory. This work can be extended to the joint channel and PHN estimation within the proposed receiver structure straightforwardly.

\begin{figure}[t]%t h b
\begin{center}
%\vspace{-4cm}
\includegraphics[width=12cm]{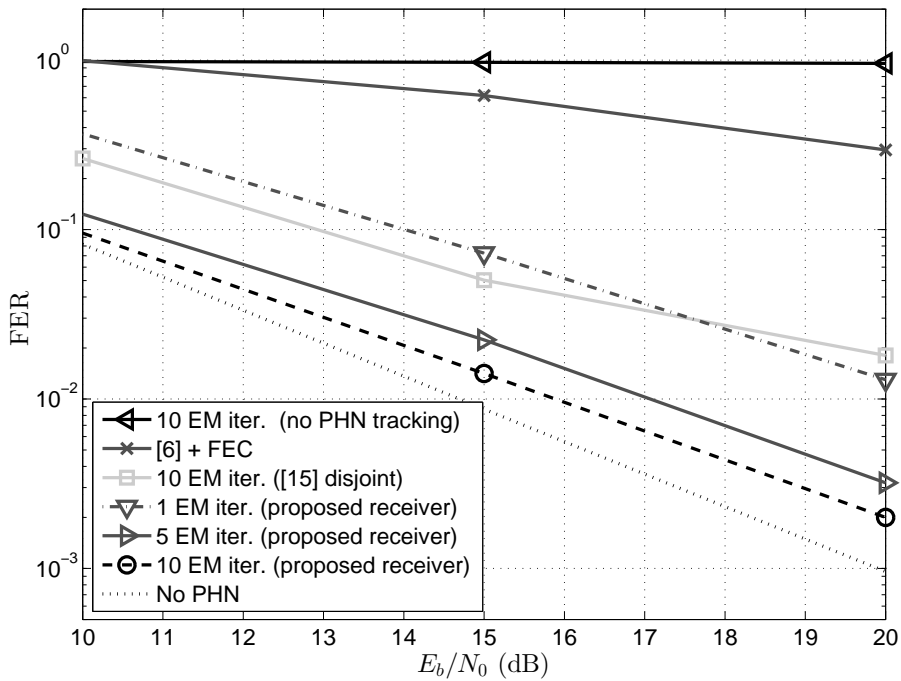}\center
%\vspace{+3.7cm}
\caption{FER of the proposed EM-based receiver ($\sigma^2_\Delta=5\cdot10^{-5}$ rad$^2$).}\label{fig:fer2EMits}
\end{center}
\vspace{-.9cm}
\end{figure}

%\begin{figure}[b]%t h b
%\begin{center}
%\includegraphics[width=8cm]{fer5e4DA14rate12.png}\center \caption{FER performance of the EM-based algorithm at several iterations where $\sigma^2_\Delta=5\cdot10^{-4}$ rad$^2$, DA initial estimation with $p_r=14$, and $R=1/2$ rate code.}\label{fig:fer5e4rate12}
%\end{center}
%\end{figure}

Fig.~\ref{fig:fer2EMits} shows the frame error-rate (FER) performance of the proposed receiver for a PHN variance of $\sigma^2_\Delta=5\cdot10^{-5}$ rad$^2$. It can be seen that the FER performance of the proposed EM receiver improves with each EM iteration. Moreover, as anticipated a MIMO receiver fails to accurately detect the received signal when no PHN tracking is applied. The results in this figure also corroborate the results in Fig.~\ref{fig:BERpr14}, showing that the proposed joint estimation and detection scheme results in significant performance gains compared to an algorithm that carries out PHN estimation and data detection separately. More specifically, a minimum performance gain of $6$ dB is observed for the same number of EM iterations. Moreover, the results in Fig.~\ref{fig:fer2EMits} show that the approach in \cite{Mehrpouyan} combined with FEC fails to achieve good frame error performance in low-to-medium SNRs. Finally, it can be observed that in this setup after $10$ EM iterations, the FER performance of the proposed receiver is only $2$ dB apart from that of perfect synchronization, i.e., ``No PHN".

 \begin{figure}[t]
\begin{center}
\vspace{-.0cm}
\includegraphics[width=12cm]{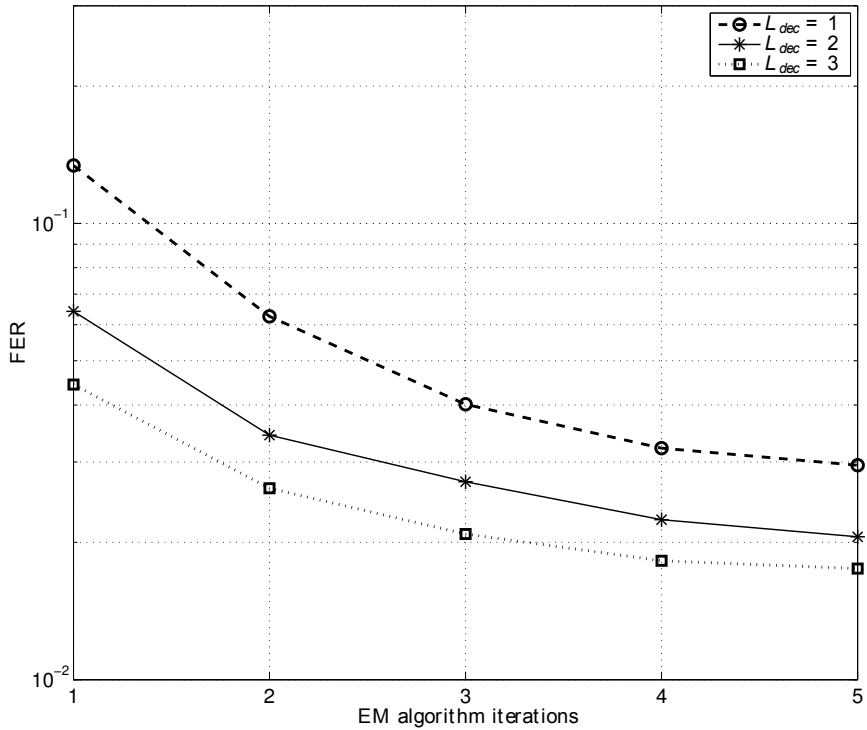}\center
%\vspace{-.3cm}
\caption{FER of the proposed EM-based receiver for several number of decoder iterations ($\sigma^2_\Delta=5\cdot10^{-4}$  rad$^2$).}\label{fig:ferLdec}
\end{center}
\vspace{-.9cm}
\end{figure}

In Fig.~\ref{fig:ferLdec} the performance of the proposed EM receiver is investigated for different number of EM and decoder iterations, $L_{dec}$. In this figure, the SNR is fixed at $E_b/N_0=20$ dB, while the PHN variance is increased to $\sigma^2_\Delta=5\cdot10^{-4}$ rad$^2$. We observe that the performance of the system improves after a few EM iterations. Moreover, the FER performance of the system can be further improved via more iterations inside of the decoder. Thus, the results in Fig.~\ref{fig:ferLdec} show that the overall performance of the MIMO system can be enhanced by increasing the number of EM or decoder iterations, which represents a clear trade-off between performance and complexity for system design.

%Fig.~\ref{fig:fer5e4rate12} shows the FER performance of the EM-based algorithm for $\sigma^2_\Delta=5\cdot10^{-4}$ rad$^2$ where data bits are encoded by a rate $R=1/2$ LDPC encoder~\cite{MacKay}. We observe that the FER performance can be improved significantly by increasing the number of EM iteration. In fact, at an FER of $5\cdot10^{-3}$ rad$^2$, the performance of the proposed receiver is only $4$ dB apart from the idealistic PHN scenario.
%
%

\vspace{-10pt}
\section{Conclusion}\label{sec:conc}

In this paper, an iterative EM-based receiver for joint PHN estimation and data detection in MIMO systems is proposed. It is demonstrated that at high SNRs, a MAP estimator can be applied to carry out the maximization step of the EM-based algorithm. However, to reduce the computational complexity of the proposed receiver, instead of a MAP estimator, an EKFS is used to carry out the maximization step of the EM algorithm. Simulation results show that the proposed receiver significantly enhances system performance. In fact, for moderate PHN variances, the overall system performance is only $1.5$ dB away from the idealistic case of perfect synchronization while applying a $7/8$ rate LDPC code. Simulation results also demonstrate that compared to receiver designs that perform PHN estimation and data detection separately, the proposed receiver results in $10$ dB and $6$ dB performance gains in terms of BER and FER, respectively. Finally, simulations indicate that the performance of the overall MIMO system can be enhanced in the presence of PHN by increasing the number of EM or decoder iterations.

\vspace{-12pt}

\bibliographystyle{IEEEtran}
\bibliography{IEEEabrv,arifbib}

\end{document}